\pdfoutput=1
\documentclass[prl,twocolumn,epsfig,amssymb,amsmath]{revtex4}

\usepackage{blkarray}

\usepackage{amssymb,amsmath,bm,epsfig,graphicx,subfigure,hyperref}

\usepackage[utf8]{inputenc}
\usepackage[vietnamese, english]{babel}
\usepackage{chngcntr}
\usepackage{xcolor}
\usepackage{braket}

\begin{document}

\title{Optically-Controlled Orbitronics on a Triangular Lattice}
\author{\foreignlanguage{vietnamese}{Võ Tiến Phong}$^{1\dagger}$}
\author{Zachariah Addison$^{1\dagger}$}
\author{Seongjin Ahn$^{2}$}
\author{Hongki Min$^2$}
\author{Ritesh Agarwal$^3$}
\author{E. J. Mele$^1$}
\email{mele@physics.upenn.edu}
\affiliation{$^1$Department of Physics and Astronomy, University of Pennsylvania, Philadelphia PA 19104}
\affiliation{$^2$Department of Physics and Astronomy, Seoul National University,  Seoul 08826, South Korea}
\affiliation{$^3$Department of Materials Science and Engineering, University of Pennsylvania, Philadelphia PA 19104}
\date{October 12, 2019}



\begin{abstract}
The propagation of electrons in an orbital multiplet dispersing on a lattice can support anomalous transport phenomena deriving from an orbitally-induced Berry curvature. In striking contrast to the related situation in graphene, we find that anomalous transport for an $L=1$ multiplet on the primitive 2D triangular lattice is activated by easily implemented on-site and optically-tunable potentials. We demonstrate this for dynamics in a Bloch band where point degeneracies carrying opposite winding numbers are generically offset in energy, allowing both an anomalous charge Hall conductance with sign selected by off-resonance coupling to circularly-polarized light and a related anomalous orbital Hall conductance activated by layer buckling.
\end{abstract}

\maketitle

Berry curvature in a band structure can manifest in anomalous responses to applied fields by inducing an anomalous velocity in the equations of motion for a wavepacket \cite{Xiao2010,Karplus1954,Sundaram1999, Thouless1982, Fang2003, Morimoto2016}. A prototype of this effect can be found on the well-studied honeycomb lattice \cite{Haldane1988,KM2005}. However, physical realizations of this model present an essential complication in practice. At half filling, the band structure has point degeneracies protected by ${\cal PT}$ symmetry that carry opposite winding numbers.  Breaking these symmetries to gap this spectrum liberates a Berry curvature into the Brillouin zone but its integrated strength vanishes unless the mass parameter also has a valley asymmetry that compensates the sign change of the winding number.  This ${\bf k}$-dependence inevitably requires site-nonlocality in the mass terms \cite{Haldane1988,KM2005} that is difficult to experimentally implement \cite{Roushan2014,Jotzu2018}. A notable work-around occurs in two-dimensional (2D) transition metal dichalcogenides where inversion symmetry is broken, and the spectrum is instead gapped by a {\it valley-symmetric} mass \cite{Xiao2012}. In this case, anomalous charge transport can be activated by a valley asymmetry in the nonequilibrium population of excited carriers produced by circularly-polarized light \cite{Xiao2012,Mak2012,Kim2014,Mak2016}. 

In this work, we consider a different approach  to engineer Berry curvatures that induce both \textit{charge} and \textit{angular momentum} anomalous Hall responses using purely local potentials in simple Bravais lattices with minimal symmetries, and propose possible experimental signatures using a representative triangular lattice. Our model is sufficiently generic that conclusions derived from it are expected to hold in many similar systems.  We are motivated by a recent work on a two-dimensional metal silicide ${\rm Cu_2Si}$ that hosts symmetry-protected line degeneracies without essential support from any sublattice symmetry and with negligible spin-orbit coupling \cite{Feng2017}, which we also assume throughout our work.  Band degeneracies in the model arise from an on-site $L=1$ orbital multiplet and are lifted by dispersion on the lattice.  Unlike the situation on the honeycomb lattice, here the winding number around the point nodes is valley-symmetric, and the net winding over the composite manifold is compensated by point degeneracies enforced elsewhere in the band structure. In this situation, regions of momentum space carrying compensating Berry curvatures are spectrally separated.  Thus, we can suppress the competing contributions of the Berry curvature to the anomalous Hall conductance (AHC) by a judicious choice of chemical potential. We demonstrate this effect on the triangular lattice by gapping out the point degeneracies that are originally protected by time-reversal symmetry via coherent coupling of the lattice to circularly-polarized light.  We find that the magnitude of the gap can be ``resonantly" enhanced by the frequency of the field. We estimate that the mass gap of a typical material on the order of 100 meV can be achieved by optical fields in the wavelength range of $0.1-1.15$ $\mu$m with experimentally-accessible intensities of $10^3-10^5$ W/$\mu$m$^2.$ A wide tunable gap means that the anomalous Hall effect in such a system should be experimentally detectable in a large range of chemical potentials.  

Next, we utilize the orbital degree of freedom to propose an anomalous \textit{orbital} Hall effect that can be activated by  layer buckling \cite{Bernevig2005, Kontani2009}. This is a transverse current to an applied field where the orbitals are polarized in the out-of-plane direction. To observe this effect, we need to break time-reversal symmetry and mirror symmetry across the lattice plane to hybridize the $L =1$ multiplets with the $L = 0$ singlet. We demonstrate this effect on the triangular lattice by calculating the anomalous orbital Hall conductance (AOHC) in the presence of a mirror-breaking perturbation. Similar phenomena should be ubiquitous in band structures which disperse an orbital multiplet, where there are degeneracies protected by mirror symmetry that can be lifted via, for instance, layer buckling. Examples of recently isolated 2D layers that host these mirror-protected line nodes include Cu$_2$Si, CuSe, and AgTe \cite{Feng2017, Gao2018, Liu2019}. With the recent surge in experimental interest in mirror-protected fermions, we expect our generic model to find  applicability in a wide number of experimental platforms. 

\begin{figure*}[htbp]
\centering
\includegraphics[width=7in]{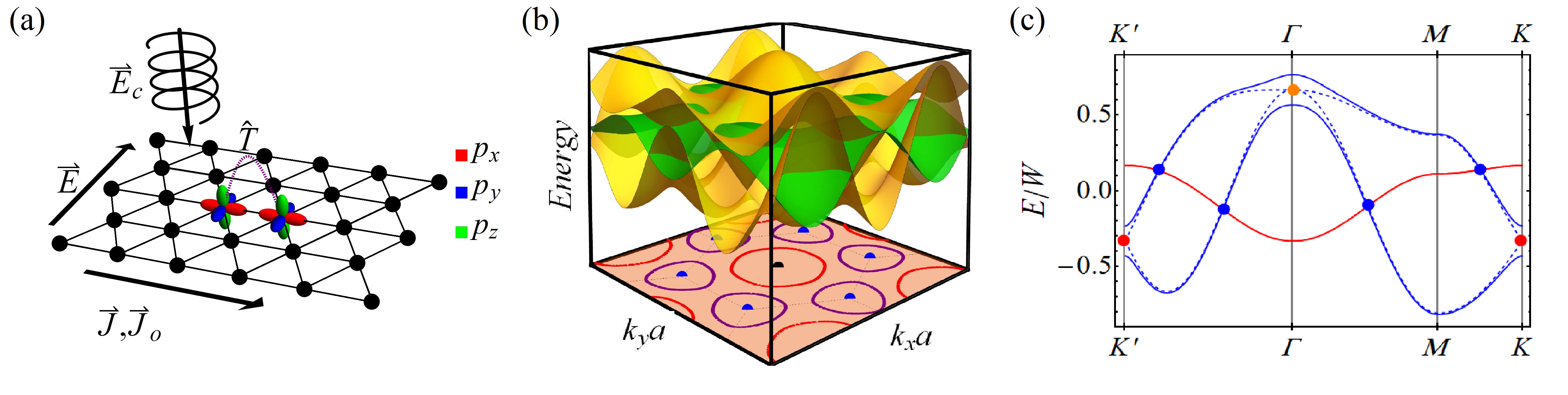} \caption{\label{latticemodel} (a) Model for the propagation of an $L=1$ orbital multiplet on a triangular lattice. Here, $\hat{T}$ is the hopping operator, $\mathbf{E}_c$ denotes a circularly-polarized optical field, and $\mathbf{J}$ and $\mathbf{J}_o$ are the charge and orbital responses to an applied transverse $\mathbf{E}$ field. (b) Energy surfaces for bands that are even (odd) under $z$-reflection shown in yellow (green). Surfaces with opposite mirror eigenvalues intersect on line nodes (projected red and purple lines). Twofold degenerate point nodes are shown as black/blue points with a quadratic contact at the zone center (black) and linear band contacts the zone corners (blue).  (c) The dispersion of the composite bands along symmetry directions in a model with ${\cal T}$-symmetry (dashed) and with a ${\cal T}$-breaking potential (solid). Blue points denote intersection of the nodal lines with the plane of the figure, and red points are twofold degeneracies pinned to high-symmetry points. Simulation parameters  are given in \cite{SI}. All energies are scaled relative to $W,$ the bandwidth at $\Gamma.$}
\end{figure*}

\begin{figure}[htbp]
\centering
\includegraphics[width=3in]{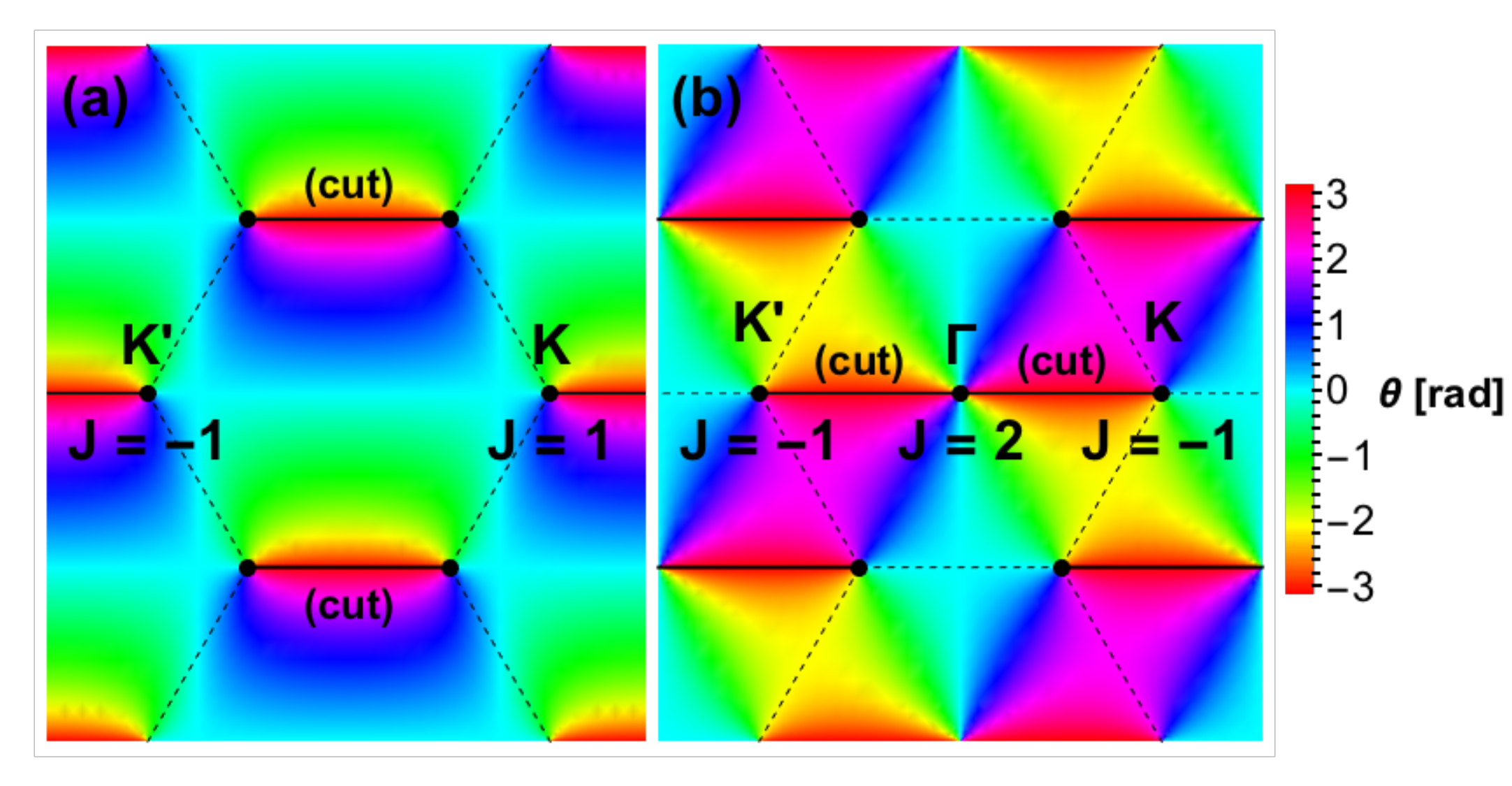} \caption{ Density plots of $\arg[d({\mathbf{k}})]$ for  tight-binding Hamiltonians of a scalar field on the honeycomb lattice (a) and for the mirror-even states on the $L=1$ triangular lattice (b). In (a), a branch cut connects point singularities at time-reversed points $K$ and $K'$. In (b), time-reversed zone-corner singularities carry the same winding number, and connect to a compensating second-order node at $\Gamma$.}\label{winding}
\end{figure}

Our lattice model derives from the propagation of an $L=1$ orbital multiplet on a triangular lattice, as illustrated in Fig. \ref{latticemodel}a. In particular, we consider a tight-binding model where each lattice site consists of three $p$ orbitals, and allow only  nearest-neighbor hoppings. The orbitals can be equivalently represented in the axial basis as $p_{+1},$ $p_0,$ and $p_{-1},$ or in the Cartesian basis as $p_x,$ $p_y,$ and $p_z.$ In either representation, the Bloch Hamiltonian at each crystal momentum $\mathbf{k} = (k_x,k_y)$ can be partitioned into these operators
\begin{equation}
\label{eq: unperturbed Hamil 1}
\hat{\mathcal{H}}(\mathbf{k}) = h_0({\bf k}) \, \hat{\mathbb{I}} + h_c({\bf k}) \, \hat l_z \cdot \hat l_z + {\bf{h}}({\bf k}) \cdot \hat{\mathbb{L}},
\end{equation}
where  $h_0(\mathbf{k})$ is a scalar coupling, $h_c(\mathbf{k})$ is a crystal field, ${\bf h}({\bf k})= \left(h_1(\mathbf{k}), h_2(\mathbf{k}) \right)$ is a vector coupling to the orbital degree of freedom, $ \hat{\mathbb{L}}= \left( [\hat{l}_x \cdot \hat{l}_x- \hat{l}_y \cdot \hat{l}_y], [\hat{l}_x \cdot \hat{l}_y+ \hat{l}_y \cdot \hat{l}_x]\right).$ The explicit forms of these $\mathbf{k}$-dependent functions are given in the Supplemental Information. The scalar term describes the average dispersion of the orbital multiplet, and the crystal field distinguishes states that are even and odd under reflection through the lattice plane. Important quantum geometry is contained in the last term of Eq. (\ref{eq: unperturbed Hamil 1}) that couples the orbital polarization to an effective ${\bf{k}}$-dependent ordering field. 

It is straightforward to verify that the Hamiltonian respects sixfold rotation symmetry of the lattice and also respects time-reversal symmetry $\mathcal{T}$. Furthermore, the in-plane subspace ($p_x,$ $p_y,$ or $p_\pm$) is decoupled from the out-of-plane subpace ($p_z$ or $p_0$). This is a consequence of mirror-reflection symmmetry about the \textit{x}-\textit{y} mirror plane that maps $(x,y,z) \mapsto (x,y,-z).$ As emphasized in Ref. \cite{Feng2017}, intersections between energy surfaces in the $z$-mirror even and odd sectors are nodal lines that are twofold degenerate in the absence of spin-orbit coupling (Fig. \ref{latticemodel}b). Because of this decoupling, we can project the Hamiltonian to just the mirror-even sector for analysis of a two-band model.  In the axial representation, the projected Hamiltonian can be written in the chiral form
\begin{equation}
\hat{\mathbb{P}}^{-1} \hat{\mathcal{H}}^\text{axi}(\mathbf{k})\hat{\mathbb{P}} = \left(h_0(\mathbf{k})+h_c(\mathbf{k}) \right) \hat{\mathbb{I}} + \hat{h}_e(\mathbf{k}),
\end{equation}
where $\hat{\mathbb{P}}$ is the mirror-even projection operator,
\begin{equation}
\label{eq: chiral}
\hat{h}_e(\mathbf{k}) = \begin{pmatrix}
0 & d(\mathbf{k}) \\
d^*(\mathbf{k}) & 0
\end{pmatrix},
\end{equation}
and $d(\mathbf{k}) = h_1(\mathbf{k}) - i h_2(\mathbf{k}),$ where $h_1(\mathbf{k})$ and $h_2(\mathbf{k})$ are defined in \cite{SI}.  Here, the $\sigma_z$ term is forbidden by the composite ${\cal T}C_{2z}$ symmetry. This feature distinguishes the primitive lattice model from its honeycomb counterpart where the ``bare" $C_{2z}$ rotation is not a symmetry of the tight binding Hamiltonian and instead is supplemented by the sublattice exchange operation $\sigma_x$.  Band degeneracies in the Hamiltonian (\ref{eq: chiral}) impose simultaneous null conditions on the real and imaginary parts of $d({\bf k})$,  which can occur only at exceptional points in two dimensions. Threefold rotational symmetry pins these points to high-symmetry momenta $\Gamma$, $K$ and $K'$ where the small groups admit two-dimensional irreducible representations (Fig. \ref{latticemodel}c). Near the $K(K')$ points, the degeneracy is lifted to linear order in momentum, while it is lifted to quadratic order at the $\Gamma$ point.

Although the linear nodes at $K$ and $K'$  are reminiscent of the situation in graphene, here its geometric character is entirely different. This is because, as mentioned above, $C_{2z}$ {\it without basis exchange} is a symmetry of the triangular lattice.  This requires the phase winding of the Bloch bands around the valley singularities to be the same. $\mathcal{T}$-symmetry requires the net winding number integrated over the full orbital manifold to vanish, and this is accomplished by a compensation from the quadratic node at $\Gamma$. Fig. \ref{winding} illustrates this point by comparing the winding of $\arg(d({\bf k}))$ for the honeycomb lattice (left), where there is a branch cut that connects the $K$ and $K'$ points, and for the $L=1$ manifold on the triangular lattice (right), where there are two branch cuts each linking a zone corner to the second-order node at $\Gamma$.

\begin{figure}[htbp]
\epsfxsize=3.6in \centerline{\epsfbox{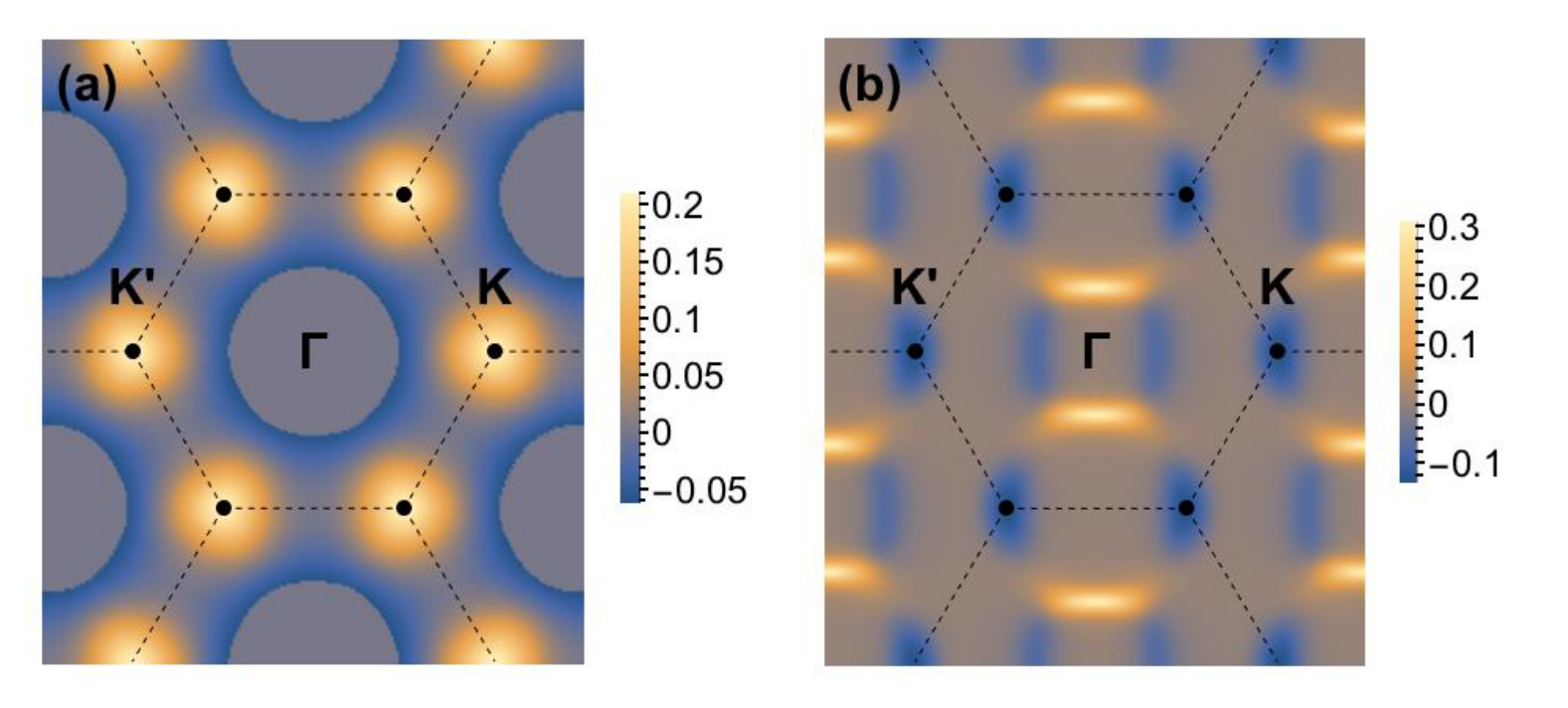}} \caption{\label{CurvatureDensities} Density plots of the occupation-weighted Berry curvature defined by the summand in Eq. (\ref{eq: Kubo}) (a) and the occupation-weighted orbital Berry curvature defined by the summand in Eq. (\ref{eq: orbital Kubo}) (b) when band degeneracies are lifted by uniform local ${\cal T}$-breaking fields. In (a), ${\cal T}$-symmetry is broken by coherent coupling to an optical field. In (b), $z$-mirror-even and odd sectors hybridize, replacing nodal lines by a quartet of linear point degeneracies. Simulation parameters  are given in \cite{SI}. The color scales are given in units of $a^2,$ where $a$ is the lattice constant.}
\end{figure}

The symmetry of the phase profile in Fig. \ref{winding} allows anomalous transport to be activated while retaining a valley-symmetric population in the presence of {\it local and spatially uniform} mass terms. Perhaps the simplest possibility is to augment the Hamiltonian of Eq. (\ref{eq: chiral}) with a ${\bf k}$-independent coupling $\varepsilon \, \sigma_z$ that breaks the degeneracy of the $m = \pm 1$ basis states, as detailed in \cite{SI}. Physical realizations include a ferromagnetic state with coupling between the magnetization and the on-site orbital moments or (as described below) coherently driving the orbital degrees of freedom with a circularly-polarized optical field.

\begin{figure*}[htbp]
\centering
\includegraphics[width=7in]{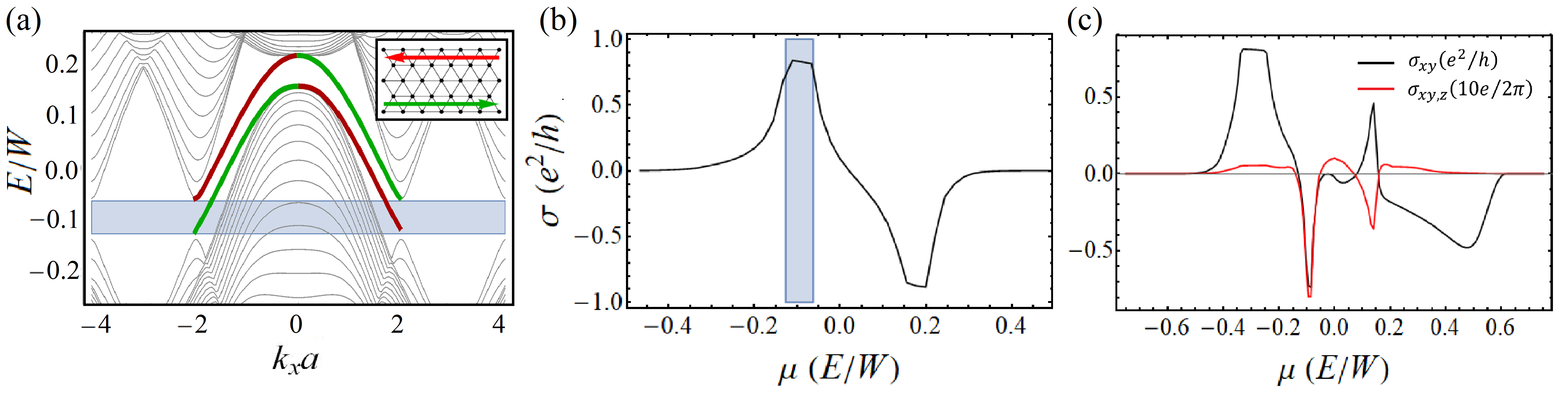} \caption{\label{Halleffects} (a) Band dispersion for a ribbon of the triangular lattice showing the projections of the bulk bands (grey) and confined edge modes inside the bulk gaps (red and green).  (Inset) The $K$ point gaps host counterpropagating states on opposite edges of the ribbon. (b) The anomalous Hall conductance  in the presence of a ${\cal T}$-breaking potential evolves from a particle-like to a hole-like response as a function of band filling $\mu$. Plateaus occur when the chemical potential is tuned within the induced gaps at the $K$ and $\Gamma$ points. (c) Breaking $z$-reflection symmetry activates an orbital Hall conductance describing a transverse angular-momentum current driven by an in-plane electric field. The orbital response is strong in two sharp spectral features where the orbital current is correlated/anticorrelated with the charge current. Simulation parameters are given in \cite{SI}.}
\end{figure*}

The effects on the band structure are shown in Fig. \ref{latticemodel}c where the band degeneracies are lifted at ${\cal O} (\varepsilon)$ at the $K(K')$ and $\Gamma$ points. Fig. \ref{CurvatureDensities}a gives a density plot of the distribution of Berry curvature in the occupied states when the chemical potential is tuned to middle of the $K$-point gaps, showing ``hot regions" near the zone-corners. In weak coupling, $\varepsilon < \Delta$, the bands overlap and the integrated Berry curvature near the zone corners is partially screened by the compensating curvature that is peaked in the higher-energy states near the $\Gamma$ point. Conversely, in strong coupling,  $\varepsilon > \Delta,$ the spectrum is fully gapped and its Chern number is zero because of an exact cancellation of these competing contributions. This latter system is adiabatically connected to a topologically-trivial system of decoupled lattice sites.  In the weak-coupling regime where the system still retains a Fermi surface, the AHC varies continuously with band filling and is given by 
\begin{equation}
\label{eq: Kubo}
\sigma _{\alpha \beta} = \frac{e^2}{\hbar} \sum_n  \int_\text{BZ} \frac{d^2 \mathbf{k}}{(2 \pi)^2} \Omega_{\alpha \beta, n} (\mathbf{k}) \Theta(\mu - \varepsilon_n(\mathbf{k}))
\end{equation}
where $\Theta(x)$ is the Heaviside step function, representing the Fermi-Dirac occupation function at zero temperature, $\mu$ is the chemical potential, the antisymmetric Berry curvature tensor for the $n^\text{th}$ band with Bloch state $|u_n({\bf k})\rangle$ and energy $\varepsilon_n(\mathbf{k})$ is given by
\begin{widetext}
\begin{equation}
\label{eq: curvature}
\Omega_{\alpha \beta,n}(\mathbf{k}) = 2 \hbar^2\sum_{m\neq n} \text{Im} \left[ \frac{\bra{u_n(\mathbf{k})}v_\alpha\ket{u_m(\mathbf{k})}\bra{u_m(\mathbf{k})}v_\beta\ket{u_n(\mathbf{k})}}{(\varepsilon_n(\mathbf{k})-\varepsilon_m(\mathbf{k}))^2}  \right],
\end{equation}
\end{widetext}
and $v_\alpha = \hbar^{-1}\partial_{k_\alpha} \mathcal{H}$ is the band velocity operator.  Fig. \ref{Halleffects}b shows the AHC as $\mu$ is swept through the spectrum. It switches from particle- to hole-like response as a function of the band filling, reflecting the proximity to the nearest sources/sinks of Berry flux; additionally, a plateau occurs when $\mu$ lies within the $K$-point gaps, where the value of the Hall conductance saturates at a value $\sigma_{xy} < e^2/h$ because of partial screening from the curvature from the higher-energy Berry sinks. In the extreme weak-coupling limit, $\varepsilon \ll \Delta$, this compensation is negligible and $\sigma_{xy}$ sharply peaks at $\sim e^2/h$ in a narrow range of $\mu$. With the chemical potential in the $K$-point gap, the AHC can also be understood in an edge-state picture (Fig. \ref{Halleffects}a) where quantized transport through edge channels is partially screened by backflow through bulk states that carry a residual curvature.
 
  Anomalous Hall response can be activated by coherently driving the system with a perpendicular circularly-polarized optical field at normal incidence which breaks $\mathcal{T}$-symmetry and lifts band degeneracies at the $\Gamma$ and $K(K')$ points. Since this field carries integer angular momentum, in lowest order, it hybridizes the mirror-even and mirror-odd bands; integrating out the latter induces an effective orbital Zeeman field $\varepsilon(\mathbf{k}) \sigma_z$  seen in the mirror-even subspace. To estimate its size, we couple the optical field to the on-site moments and calculate the mass term, as derived in \cite{SI}, to be
\begin{equation}
\label{eq: field-induced gap}
\varepsilon(\mathbf{k}) = \frac{e^2E_0^2 \mathfrak{p}^2 \hbar  \omega }{2(h_c(\mathbf{k})^2-\hbar^2 \omega^2)},
\end{equation}
which is second order in the driving field $E_0$, linear in  the driving frequency $\omega$,  and controlled by the strength of the interorbital matrix element $\mathfrak{p}$ .  This  mass can be resonantly-tuned by adjusting the driving frequency through the crystal field scale. To estimate the size of this effect, we take representative parameters for a typical material, $h_c \approx 1$ eV and $\mathfrak{p} = 1 \text{ }{\rm \AA}$, achieving a mass scale $\sim 100$ meV in the wavelength range $ 0.1-1.15$ $\mu$m (resonance is at $1.24$ $\mu$m) requires peak intensities in the range $10^3-10^5$ W/$\mu$m$^2,$ which is accessible to currently available sources.

The orbital degree of freedom also allows the possibility of an {\it angular momentum} current where an anomalous flow of orbital angular momentum, with or without charge, is directed perpendicular to an applied in-plane electric field \cite{Bernevig2005, Kontani2009}.  A natural choice for the angular momentum current operator is $j_{ \beta}^{(\alpha)} = \frac{\hbar}{2}\{l_\alpha, v_\beta\}$. However, because the angular momentum operators $l_\alpha$ do not commute with the Hamiltonian, such a current operator does not satisfy a continuity equation. Instead, in the regime where one is probing low-frequency dynamics with a period much larger than the interband dephasing time, we can use a band-projected version where $l_\alpha$ is replaced by $\sum_n {\cal P}_{{\bf k},n} l_\alpha{\cal P}_{{\bf k},n},$ and ${\cal P}_{{\bf k},n} = \ket{u_{n}(\mathbf{k})}\bra{u_n(\mathbf{k})}$ is the projection operator. This operator projects the angular momentum operators onto the diagonal elements of the density matrix, and clearly commutes with the Hamiltonian. In this low-frequency regime, we can write the angular momentum current operator as \cite{Murakami2003}
\begin{eqnarray}
 j_{ \beta}^{(\alpha)} = \frac{\hbar}{2} \sum_n \left\{{\cal P}_{{\bf k},n} l_\alpha {\cal P}_{{\bf k},n}, v_\beta \right\}
\end{eqnarray}
to describe a current flowing in the $\beta$-direction with angular momentum polarized along the $\alpha$-direction. The anomalous orbital transport coefficient derived from $j_{\alpha \beta}$ is purely transverse, leading to  $J^{(z)}_{\alpha} =    \sigma^{(z)}_{\alpha \beta} E_\beta$, where $\sigma^{(z)}_{\alpha \beta }$ contains an angular-momentum-weighted curvature 
\begin{equation}
\label{eq: orbital Kubo}
\sigma^{(z)}_{\alpha \beta} =  e \sum_n \int_\text{BZ} \frac{d^2 \mathbf{k}}{(2 \pi)^2} \Omega^{(z)}_{\alpha \beta,n}(\mathbf{k}) \Theta (\mu - \varepsilon_n({\bf k})),
\end{equation}
where the angular-momentum-weighted curvature is \cite{Go2018}
\begin{widetext}
\begin{equation}
\label{eq: orbital curvature}
\Omega^{(z)}_{\alpha \beta,n}(\mathbf{k}) = 2\hbar \sum_{m\neq n} \text{Im} \left[ \frac{\bra{u_n(\mathbf{k})} j_{\alpha}^{(z)}\ket{u_m(\mathbf{k})}\bra{u_m(\mathbf{k})}v_\beta\ket{u_n(\mathbf{k})}}{(\varepsilon_n(\mathbf{k})-\varepsilon_m(\mathbf{k}))^2}  \right].
\end{equation}
\end{widetext}
We note that the orbital curvature defined in Eq. (\ref{eq: orbital curvature}) is reminiscent of the charge curvature in Eq. (\ref{eq: curvature}) with the charge current operator replaced by the angular momentum current operator. For the special case of a two-band model where the band-projected angular momenta $\langle l_z \rangle$  exactly cancel, the AOHC in Eq. \ref{eq: orbital Kubo} vanishes. In our model, it is activated by breaking $z$-mirror symmetry which hybridizes the in-plane and the out-of-plane orbital polarizations, as detailed in \cite{SI}.  This can be accomplished via layer buckling. Fig. \ref{CurvatureDensities}b shows the distribution of orbital Berry curvature produced by a potential which retains $y$-mirror symmetry but breaks the horizontal mirror symmetry by mixing $x$- and $z$-orbital polarizations. This lifts the line-node degeneracy except for a quartet of exceptional band contact points.  The orbital curvature is largest at momenta where in-plane and out-of-plane polarizations are optimally mixed. Fig. \ref{Halleffects}c shows the charge and orbital Hall conductances calculated in this three-band model as a function of $\mu$. The broken symmetry activates the AOHC seen in two sharply-defined spectral features where the in-plane and out-of-plane degrees of freedom are most strongly mixed. These modes describe anomalous transport of charge and angular momentum  that are correlated (anticorrelated) in the lower (upper) bands. Interestingly, we find that in the strong-coupling limit where the  spectrum is fully gapped, the AHC vanishes but the residual AOHC  retains nonzero plateau representing a pure flow of angular momentum with no concomitant flow of charge. Experimentally, the orbital Hall effect is established in a two-dimensional system by measuring the out-of-plane angular momentum polarization of the transverse current to an applied field.


Related phenomena can be expected in other situations where propagation on a lattice disperses an orbital degree of freedom \cite{Tokura2000}. Although our study is motivated by the band topology in ${\rm Cu_2Si}$ \cite{Feng2017}, this material is not optimal for this application because the relevant Si-$3p$-derived bands overlap with Cu-$3d$ states, obscuring some of the most interesting singularities in the active band manifold. One expects that this obstacle can be circumvented by a judicious choice of cations in related materials. Recently, 2D lattices of CuSe and AgTe have been successfully fabricated and characterized. The lattice structure of these materials has a triangular sublattice and has been shown to host fermion line nodes that are protected by mirror symmetry in the absence of spin-orbit coupling \cite{Gao2018, Liu2019}. Importantly, since the orbital connection does not rely on spin-orbit coupling, our approach can immediately be used to support topological transport in systems containing only light elements.  We note that previous theoretical work along these lines considered a 2D variant of this model where the in-plane orbital degrees of freedom are instead coupled on the two-site basis of a {\it honeycomb} lattice, showing that it can realize an orbital analog of the quantum anomalous Hall effect \cite{Wu2008} and even realizations on optical lattices \cite{Wu2007}. The use of an on-site vector degree of freedom along with broken ${\cal T}$ symmetry lends itself naturally to topological {\it mechanical} systems in a driven state designed to break reciprocity \cite{Nash2015}. We also note that the model derived here for describing the lattice propagation of an integer-quantized orbital multiplet is (absent the crystal field splitting) a 2D variant of a 3D model that possesses point nodes for $J=1$ lattice fermions \cite{Tang2017}. Spin-orbit coupling on the triangular lattice can also lead to other topological phases such as the quantum spin Hall effect \cite{Liang2016}.

Symmetry and response function analysis  in terms of Berry curvatures (ZA and EJM) was supported by the Department of Energy under grant DE-FG02-84ER45118. VTP acknowledges support from the NSF Graduate Research Fellowships Program and the P.D. Soros Fellowship for New Americans, and additional support from the Department of Energy. S.A. and H.M. were supported by the NRF grant funded by the Korea government (MSIT) (No. 2018R1A2B6007837) and Creative-Pioneering Researchers Program through Seoul National University (SNU). RA is supported by a grant from the US Army Research Office W911NF-17-1-0436.

$^\dagger$VTP and ZA contributed equally to this work.

\bibliography{Orbitronics_PRL_Resubmission}

\begin{thebibliography}{29}
\expandafter\ifx\csname natexlab\endcsname\relax\def\natexlab#1{#1}\fi
\expandafter\ifx\csname bibnamefont\endcsname\relax
  \def\bibnamefont#1{#1}\fi
\expandafter\ifx\csname bibfnamefont\endcsname\relax
  \def\bibfnamefont#1{#1}\fi
\expandafter\ifx\csname citenamefont\endcsname\relax
  \def\citenamefont#1{#1}\fi
\expandafter\ifx\csname url\endcsname\relax
  \def\url#1{\texttt{#1}}\fi
\expandafter\ifx\csname urlprefix\endcsname\relax\def\urlprefix{URL }\fi
\providecommand{\bibinfo}[2]{#2}
\providecommand{\eprint}[2][]{\url{#2}}

\bibitem[{\citenamefont{Xiao et~al.}(2010)\citenamefont{Xiao, Chang, and
  Niu}}]{Xiao2010}
\bibinfo{author}{\bibfnamefont{D.}~\bibnamefont{Xiao}},
  \bibinfo{author}{\bibfnamefont{M.-C.} \bibnamefont{Chang}}, \bibnamefont{and}
  \bibinfo{author}{\bibfnamefont{Q.}~\bibnamefont{Niu}}, \bibinfo{journal}{Rev.
  Mod. Phys.} \textbf{\bibinfo{volume}{82}}, \bibinfo{pages}{1959}
  (\bibinfo{year}{2010}).

\bibitem[{\citenamefont{Karplus and Luttinger}(1954)}]{Karplus1954}
\bibinfo{author}{\bibfnamefont{R.}~\bibnamefont{Karplus}} \bibnamefont{and}
  \bibinfo{author}{\bibfnamefont{J.~M.} \bibnamefont{Luttinger}},
  \bibinfo{journal}{Phys. Rev.} \textbf{\bibinfo{volume}{95}},
  \bibinfo{pages}{1154} (\bibinfo{year}{1954}).

\bibitem[{\citenamefont{Sundaram and Niu}(1999)}]{Sundaram1999}
\bibinfo{author}{\bibfnamefont{G.}~\bibnamefont{Sundaram}} \bibnamefont{and}
  \bibinfo{author}{\bibfnamefont{Q.}~\bibnamefont{Niu}},
  \bibinfo{journal}{Phys. Rev. B} \textbf{\bibinfo{volume}{59}},
  \bibinfo{pages}{14915} (\bibinfo{year}{1999}).

\bibitem[{\citenamefont{Thouless et~al.}(1982)\citenamefont{Thouless, Kohmoto,
  Nightingale, and den Nijs}}]{Thouless1982}
\bibinfo{author}{\bibfnamefont{D.~J.} \bibnamefont{Thouless}},
  \bibinfo{author}{\bibfnamefont{M.}~\bibnamefont{Kohmoto}},
  \bibinfo{author}{\bibfnamefont{M.~P.} \bibnamefont{Nightingale}},
  \bibnamefont{and} \bibinfo{author}{\bibfnamefont{M.}~\bibnamefont{den Nijs}},
  \bibinfo{journal}{Phys. Rev. Lett.} \textbf{\bibinfo{volume}{49}},
  \bibinfo{pages}{405} (\bibinfo{year}{1982}).

\bibitem[{\citenamefont{Fang et~al.}(2003)\citenamefont{Fang, Nagaosa,
  Takahashi, Asamitsu, Mathieu, Ogasawara, Yamada, Kawasaki, Tokura, and
  Terakura}}]{Fang2003}
\bibinfo{author}{\bibfnamefont{Z.}~\bibnamefont{Fang}},
  \bibinfo{author}{\bibfnamefont{N.}~\bibnamefont{Nagaosa}},
  \bibinfo{author}{\bibfnamefont{K.~S.} \bibnamefont{Takahashi}},
  \bibinfo{author}{\bibfnamefont{A.}~\bibnamefont{Asamitsu}},
  \bibinfo{author}{\bibfnamefont{R.}~\bibnamefont{Mathieu}},
  \bibinfo{author}{\bibfnamefont{T.}~\bibnamefont{Ogasawara}},
  \bibinfo{author}{\bibfnamefont{H.}~\bibnamefont{Yamada}},
  \bibinfo{author}{\bibfnamefont{M.}~\bibnamefont{Kawasaki}},
  \bibinfo{author}{\bibfnamefont{Y.}~\bibnamefont{Tokura}}, \bibnamefont{and}
  \bibinfo{author}{\bibfnamefont{K.}~\bibnamefont{Terakura}},
  \textbf{\bibinfo{volume}{302}}, \bibinfo{pages}{92} (\bibinfo{year}{2003}),
  ISSN \bibinfo{issn}{0036-8075}.

\bibitem[{\citenamefont{Morimoto and Nagaosa}(2016)}]{Morimoto2016}
\bibinfo{author}{\bibfnamefont{T.}~\bibnamefont{Morimoto}} \bibnamefont{and}
  \bibinfo{author}{\bibfnamefont{N.}~\bibnamefont{Nagaosa}},
  \bibinfo{journal}{Science Advances} \textbf{\bibinfo{volume}{2}}
  (\bibinfo{year}{2016}).

\bibitem[{\citenamefont{Haldane}(1988)}]{Haldane1988}
\bibinfo{author}{\bibfnamefont{F.~D.~M.} \bibnamefont{Haldane}},
  \bibinfo{journal}{Phys. Rev. Lett.} \textbf{\bibinfo{volume}{61}},
  \bibinfo{pages}{2015} (\bibinfo{year}{1988}).

\bibitem[{\citenamefont{Kane and Mele}(2005)}]{KM2005}
\bibinfo{author}{\bibfnamefont{C.~L.} \bibnamefont{Kane}} \bibnamefont{and}
  \bibinfo{author}{\bibfnamefont{E.~J.} \bibnamefont{Mele}},
  \bibinfo{journal}{Phys. Rev. Lett.} \textbf{\bibinfo{volume}{95}},
  \bibinfo{pages}{226801} (\bibinfo{year}{2005}).

\bibitem[{\citenamefont{Roushan et~al.}(2014)\citenamefont{Roushan, Neill,
  Chen, Kolodrubetz, Quintana, Leung, Fang, Barends, Campbell, Chen
  et~al.}}]{Roushan2014}
\bibinfo{author}{\bibfnamefont{P.}~\bibnamefont{Roushan}},
  \bibinfo{author}{\bibfnamefont{C.}~\bibnamefont{Neill}},
  \bibinfo{author}{\bibfnamefont{Y.}~\bibnamefont{Chen}},
  \bibinfo{author}{\bibfnamefont{M.}~\bibnamefont{Kolodrubetz}},
  \bibinfo{author}{\bibfnamefont{C.}~\bibnamefont{Quintana}},
  \bibinfo{author}{\bibfnamefont{N.}~\bibnamefont{Leung}},
  \bibinfo{author}{\bibfnamefont{M.}~\bibnamefont{Fang}},
  \bibinfo{author}{\bibfnamefont{R.}~\bibnamefont{Barends}},
  \bibinfo{author}{\bibfnamefont{B.}~\bibnamefont{Campbell}},
  \bibinfo{author}{\bibfnamefont{Z.}~\bibnamefont{Chen}}, \bibnamefont{et~al.},
  \bibinfo{journal}{Nature} \textbf{\bibinfo{volume}{515}},
  \bibinfo{pages}{241} (\bibinfo{year}{2014}).

\bibitem[{\citenamefont{Jotzu et~al.}(2014)\citenamefont{Jotzu, Messer,
  Desbuquois, Lebrat, Uehlinger, Greif, and Esslinger}}]{Jotzu2018}
\bibinfo{author}{\bibfnamefont{G.}~\bibnamefont{Jotzu}},
  \bibinfo{author}{\bibfnamefont{M.}~\bibnamefont{Messer}},
  \bibinfo{author}{\bibfnamefont{R.}~\bibnamefont{Desbuquois}},
  \bibinfo{author}{\bibfnamefont{M.}~\bibnamefont{Lebrat}},
  \bibinfo{author}{\bibfnamefont{T.}~\bibnamefont{Uehlinger}},
  \bibinfo{author}{\bibfnamefont{D.}~\bibnamefont{Greif}}, \bibnamefont{and}
  \bibinfo{author}{\bibfnamefont{T.}~\bibnamefont{Esslinger}},
  \bibinfo{journal}{Nature} \textbf{\bibinfo{volume}{515}},
  \bibinfo{pages}{237} (\bibinfo{year}{2014}).

\bibitem[{\citenamefont{Xiao et~al.}(2012)\citenamefont{Xiao, Liu, Feng, Xu,
  and Yao}}]{Xiao2012}
\bibinfo{author}{\bibfnamefont{D.}~\bibnamefont{Xiao}},
  \bibinfo{author}{\bibfnamefont{G.-B.} \bibnamefont{Liu}},
  \bibinfo{author}{\bibfnamefont{W.}~\bibnamefont{Feng}},
  \bibinfo{author}{\bibfnamefont{X.}~\bibnamefont{Xu}}, \bibnamefont{and}
  \bibinfo{author}{\bibfnamefont{W.}~\bibnamefont{Yao}},
  \bibinfo{journal}{Phys. Rev. Lett.} \textbf{\bibinfo{volume}{108}},
  \bibinfo{pages}{196802} (\bibinfo{year}{2012}).

\bibitem[{\citenamefont{Mak et~al.}(2012)\citenamefont{Mak, He, Shan, and
  Heinz}}]{Mak2012}
\bibinfo{author}{\bibfnamefont{K.~F.} \bibnamefont{Mak}},
  \bibinfo{author}{\bibfnamefont{K.}~\bibnamefont{He}},
  \bibinfo{author}{\bibfnamefont{J.}~\bibnamefont{Shan}}, \bibnamefont{and}
  \bibinfo{author}{\bibfnamefont{T.~F.} \bibnamefont{Heinz}},
  \bibinfo{journal}{Nat. Nanotechnology} \textbf{\bibinfo{volume}{7}},
  \bibinfo{pages}{494} (\bibinfo{year}{2012}).

\bibitem[{\citenamefont{Kim et~al.}(2014)\citenamefont{Kim, Hong, Jin, Shi,
  Chang, Chiu, Li, and Wang}}]{Kim2014}
\bibinfo{author}{\bibfnamefont{J.}~\bibnamefont{Kim}},
  \bibinfo{author}{\bibfnamefont{X.}~\bibnamefont{Hong}},
  \bibinfo{author}{\bibfnamefont{C.}~\bibnamefont{Jin}},
  \bibinfo{author}{\bibfnamefont{S.-F.} \bibnamefont{Shi}},
  \bibinfo{author}{\bibfnamefont{C.-Y.~S.} \bibnamefont{Chang}},
  \bibinfo{author}{\bibfnamefont{M.-H.} \bibnamefont{Chiu}},
  \bibinfo{author}{\bibfnamefont{L.-J.} \bibnamefont{Li}}, \bibnamefont{and}
  \bibinfo{author}{\bibfnamefont{F.}~\bibnamefont{Wang}},
  \bibinfo{journal}{Science} \textbf{\bibinfo{volume}{346}},
  \bibinfo{pages}{1205} (\bibinfo{year}{2014}).

\bibitem[{\citenamefont{Mak and Shan}(2016)}]{Mak2016}
\bibinfo{author}{\bibfnamefont{K.~F.} \bibnamefont{Mak}} \bibnamefont{and}
  \bibinfo{author}{\bibfnamefont{J.}~\bibnamefont{Shan}},
  \bibinfo{journal}{Nat. Photonics} \textbf{\bibinfo{volume}{10}},
  \bibinfo{pages}{216} (\bibinfo{year}{2016}).

\bibitem[{\citenamefont{Feng et~al.}(2017)\citenamefont{Feng, Fu, Kasamatsu,
  Ito, Cheng, Liu, Feng, Wu, Mahatha, Sheverdyaeva et~al.}}]{Feng2017}
\bibinfo{author}{\bibfnamefont{B.}~\bibnamefont{Feng}},
  \bibinfo{author}{\bibfnamefont{B.}~\bibnamefont{Fu}},
  \bibinfo{author}{\bibfnamefont{S.}~\bibnamefont{Kasamatsu}},
  \bibinfo{author}{\bibfnamefont{S.}~\bibnamefont{Ito}},
  \bibinfo{author}{\bibfnamefont{P.}~\bibnamefont{Cheng}},
  \bibinfo{author}{\bibfnamefont{C.-C.} \bibnamefont{Liu}},
  \bibinfo{author}{\bibfnamefont{Y.}~\bibnamefont{Feng}},
  \bibinfo{author}{\bibfnamefont{S.}~\bibnamefont{Wu}},
  \bibinfo{author}{\bibfnamefont{S.~K.} \bibnamefont{Mahatha}},
  \bibinfo{author}{\bibfnamefont{P.}~\bibnamefont{Sheverdyaeva}},
  \bibnamefont{et~al.}, \bibinfo{journal}{Nat. Commun.}
  \textbf{\bibinfo{volume}{8}}, \bibinfo{pages}{1007} (\bibinfo{year}{2017}).

\bibitem[{\citenamefont{Bernevig et~al.}(2005)\citenamefont{Bernevig, Hughes,
  and Zhang}}]{Bernevig2005}
\bibinfo{author}{\bibfnamefont{B.~A.} \bibnamefont{Bernevig}},
  \bibinfo{author}{\bibfnamefont{T.~L.} \bibnamefont{Hughes}},
  \bibnamefont{and} \bibinfo{author}{\bibfnamefont{S.-C.} \bibnamefont{Zhang}},
  \bibinfo{journal}{Phys. Rev. Lett.} \textbf{\bibinfo{volume}{95}},
  \bibinfo{pages}{066601} (\bibinfo{year}{2005}).

\bibitem[{\citenamefont{Kontani et~al.}(2009)\citenamefont{Kontani, Tanaka,
  Hirashima, Yamada, and Inoue}}]{Kontani2009}
\bibinfo{author}{\bibfnamefont{H.}~\bibnamefont{Kontani}},
  \bibinfo{author}{\bibfnamefont{T.}~\bibnamefont{Tanaka}},
  \bibinfo{author}{\bibfnamefont{D.~S.} \bibnamefont{Hirashima}},
  \bibinfo{author}{\bibfnamefont{K.}~\bibnamefont{Yamada}}, \bibnamefont{and}
  \bibinfo{author}{\bibfnamefont{J.}~\bibnamefont{Inoue}},
  \bibinfo{journal}{Phys. Rev. Lett.} \textbf{\bibinfo{volume}{102}},
  \bibinfo{pages}{016601} (\bibinfo{year}{2009}).

\bibitem[{\citenamefont{Gao et~al.}(2018)\citenamefont{Gao, Sun, Lu, Li, Qian,
  Zhang, Zhang, Qian, Ding, Lin et~al.}}]{Gao2018}
\bibinfo{author}{\bibfnamefont{L.}~\bibnamefont{Gao}},
  \bibinfo{author}{\bibfnamefont{J.-T.} \bibnamefont{Sun}},
  \bibinfo{author}{\bibfnamefont{J.-C.} \bibnamefont{Lu}},
  \bibinfo{author}{\bibfnamefont{H.}~\bibnamefont{Li}},
  \bibinfo{author}{\bibfnamefont{K.}~\bibnamefont{Qian}},
  \bibinfo{author}{\bibfnamefont{S.}~\bibnamefont{Zhang}},
  \bibinfo{author}{\bibfnamefont{Y.-Y.} \bibnamefont{Zhang}},
  \bibinfo{author}{\bibfnamefont{T.}~\bibnamefont{Qian}},
  \bibinfo{author}{\bibfnamefont{H.}~\bibnamefont{Ding}},
  \bibinfo{author}{\bibfnamefont{X.}~\bibnamefont{Lin}}, \bibnamefont{et~al.},
  \bibinfo{journal}{Advanced Materials} \textbf{\bibinfo{volume}{30}},
  \bibinfo{pages}{1707055} (\bibinfo{year}{2018}).

\bibitem[{\citenamefont{Liu et~al.}(2019)\citenamefont{Liu, Liu, Miao, Xue,
  Zhang, Liu, Huang, Zhu, Meng, Guo et~al.}}]{Liu2019}
\bibinfo{author}{\bibfnamefont{B.}~\bibnamefont{Liu}},
  \bibinfo{author}{\bibfnamefont{J.}~\bibnamefont{Liu}},
  \bibinfo{author}{\bibfnamefont{G.}~\bibnamefont{Miao}},
  \bibinfo{author}{\bibfnamefont{S.}~\bibnamefont{Xue}},
  \bibinfo{author}{\bibfnamefont{S.}~\bibnamefont{Zhang}},
  \bibinfo{author}{\bibfnamefont{L.}~\bibnamefont{Liu}},
  \bibinfo{author}{\bibfnamefont{X.}~\bibnamefont{Huang}},
  \bibinfo{author}{\bibfnamefont{X.}~\bibnamefont{Zhu}},
  \bibinfo{author}{\bibfnamefont{S.}~\bibnamefont{Meng}},
  \bibinfo{author}{\bibfnamefont{J.}~\bibnamefont{Guo}}, \bibnamefont{et~al.},
  \bibinfo{journal}{arXiv preprint arXiv:1901.06284}  (\bibinfo{year}{2019}).

\bibitem[{SI()}]{SI}
\bibinfo{note}{See Supplemental Information for a derivation of the
  Hamiltonian, a discussion of the winding numbers around the degenerate
  points, details about the simulation of the Hall responses, and a derivation
  of the optical coupling.}

\bibitem[{\citenamefont{Murakami et~al.}(2003)\citenamefont{Murakami, Nagaosa,
  and Zhang}}]{Murakami2003}
\bibinfo{author}{\bibfnamefont{S.}~\bibnamefont{Murakami}},
  \bibinfo{author}{\bibfnamefont{N.}~\bibnamefont{Nagaosa}}, \bibnamefont{and}
  \bibinfo{author}{\bibfnamefont{S.-C.} \bibnamefont{Zhang}},
  \bibinfo{journal}{Science} \textbf{\bibinfo{volume}{301}},
  \bibinfo{pages}{1348} (\bibinfo{year}{2003}).

\bibitem[{\citenamefont{Go et~al.}(2018)\citenamefont{Go, Jo, Kim, and
  Lee}}]{Go2018}
\bibinfo{author}{\bibfnamefont{D.}~\bibnamefont{Go}},
  \bibinfo{author}{\bibfnamefont{D.}~\bibnamefont{Jo}},
  \bibinfo{author}{\bibfnamefont{C.}~\bibnamefont{Kim}}, \bibnamefont{and}
  \bibinfo{author}{\bibfnamefont{H.-W.} \bibnamefont{Lee}},
  \bibinfo{journal}{Phys. Rev. Lett.} \textbf{\bibinfo{volume}{121}},
  \bibinfo{pages}{086602} (\bibinfo{year}{2018}).

\bibitem[{\citenamefont{Tokura and Nagaosa}(2000)}]{Tokura2000}
\bibinfo{author}{\bibfnamefont{Y.}~\bibnamefont{Tokura}} \bibnamefont{and}
  \bibinfo{author}{\bibfnamefont{N.}~\bibnamefont{Nagaosa}},
  \bibinfo{journal}{Science} \textbf{\bibinfo{volume}{288}},
  \bibinfo{pages}{462} (\bibinfo{year}{2000}).

\bibitem[{\citenamefont{Wu}(2008)}]{Wu2008}
\bibinfo{author}{\bibfnamefont{C.}~\bibnamefont{Wu}}, \bibinfo{journal}{Phys.
  Rev. Lett.} \textbf{\bibinfo{volume}{101}}, \bibinfo{pages}{186807}
  (\bibinfo{year}{2008}).

\bibitem[{\citenamefont{Wu et~al.}(2007)\citenamefont{Wu, Bergman, Balents, and
  Das~Sarma}}]{Wu2007}
\bibinfo{author}{\bibfnamefont{C.}~\bibnamefont{Wu}},
  \bibinfo{author}{\bibfnamefont{D.}~\bibnamefont{Bergman}},
  \bibinfo{author}{\bibfnamefont{L.}~\bibnamefont{Balents}}, \bibnamefont{and}
  \bibinfo{author}{\bibfnamefont{S.}~\bibnamefont{Das~Sarma}},
  \bibinfo{journal}{Phys. Rev. Lett.} \textbf{\bibinfo{volume}{99}},
  \bibinfo{pages}{070401} (\bibinfo{year}{2007}).

\bibitem[{\citenamefont{Nash et~al.}(2015)\citenamefont{Nash, Kleckner, Read,
  Vitelli, Turner, and Irvine}}]{Nash2015}
\bibinfo{author}{\bibfnamefont{L.~M.} \bibnamefont{Nash}},
  \bibinfo{author}{\bibfnamefont{D.}~\bibnamefont{Kleckner}},
  \bibinfo{author}{\bibfnamefont{A.}~\bibnamefont{Read}},
  \bibinfo{author}{\bibfnamefont{V.}~\bibnamefont{Vitelli}},
  \bibinfo{author}{\bibfnamefont{A.~M.} \bibnamefont{Turner}},
  \bibnamefont{and} \bibinfo{author}{\bibfnamefont{W.~T.~M.}
  \bibnamefont{Irvine}}, \bibinfo{journal}{Proc. Natl. Acad. Sci. U. S. A.}
  \textbf{\bibinfo{volume}{112}}, \bibinfo{pages}{14495}
  (\bibinfo{year}{2015}).

\bibitem[{\citenamefont{Tang et~al.}(2017)\citenamefont{Tang, Zhou, and
  Zhang}}]{Tang2017}
\bibinfo{author}{\bibfnamefont{P.}~\bibnamefont{Tang}},
  \bibinfo{author}{\bibfnamefont{Q.}~\bibnamefont{Zhou}}, \bibnamefont{and}
  \bibinfo{author}{\bibfnamefont{S.-C.} \bibnamefont{Zhang}},
  \bibinfo{journal}{Phys. Rev. Lett.} \textbf{\bibinfo{volume}{119}},
  \bibinfo{pages}{206402} (\bibinfo{year}{2017}).

\bibitem[{\citenamefont{Liang et~al.}(2016)\citenamefont{Liang, Yu, Zhou, and
  Hu}}]{Liang2016}
\bibinfo{author}{\bibfnamefont{Q.-F.} \bibnamefont{Liang}},
  \bibinfo{author}{\bibfnamefont{R.}~\bibnamefont{Yu}},
  \bibinfo{author}{\bibfnamefont{J.}~\bibnamefont{Zhou}}, \bibnamefont{and}
  \bibinfo{author}{\bibfnamefont{X.}~\bibnamefont{Hu}}, \bibinfo{journal}{Phys.
  Rev. B} \textbf{\bibinfo{volume}{93}}, \bibinfo{pages}{035135}
  (\bibinfo{year}{2016}).

\bibitem[{\citenamefont{Novi\ifmmode~\check{c}\else \v{c}\fi{}enko
  et~al.}(2017)\citenamefont{Novi\ifmmode~\check{c}\else \v{c}\fi{}enko,
  Anisimovas, and Juzeli\ifmmode~\bar{u}\else \={u}\fi{}nas}}]{Novi2017}
\bibinfo{author}{\bibfnamefont{V.}~\bibnamefont{Novi\ifmmode~\check{c}\else
  \v{c}\fi{}enko}},
  \bibinfo{author}{\bibfnamefont{E.}~\bibnamefont{Anisimovas}},
  \bibnamefont{and}
  \bibinfo{author}{\bibfnamefont{G.}~\bibnamefont{Juzeli\ifmmode~\bar{u}\else
  \={u}\fi{}nas}}, \bibinfo{journal}{Phys. Rev. A}
  \textbf{\bibinfo{volume}{95}}, \bibinfo{pages}{023615}
  (\bibinfo{year}{2017}).

\end{thebibliography}

\pagebreak
\appendix

\onecolumngrid

\setcounter{equation}{0}
\setcounter{figure}{0}

\renewcommand{\theequation}{S\arabic{equation}}
\renewcommand{\thefigure}{S\arabic{figure}}

\renewcommand{\bibnumfmt}[1]{[#1]}
\renewcommand{\citenumfont}[1]{#1}

\section{A. Hamiltonian in Momentum Space}

Our lattice model derives from the propagation of an $L=1$ orbital multiplet on a triangular lattice, as illustrated in Fig. 1a. In particular, we consider a tight-binding model where each lattice site consists of three localized $p$ orbitals, and allow hoppings only between nearest-neighbor sites. The orbitals can be equivalently represented in the axial basis as $p_{+1},$ $p_0,$ and $p_{-1},$ or in the Cartesian basis as $p_x,$ $p_y,$ and $p_z.$ In the axial representation, the angular momentum operators can be written in the conventional $\hat{l}_z$-diagonal basis as
\begin{equation}
\hat{l}_x^\text{axi} = \frac{1}{\sqrt{2}} \begin{pmatrix}
0 & 1 & 0 \\
1 & 0 & 1 \\
0 & 1 & 0
\end{pmatrix}, \quad \hat{l}_y^\text{axi} = \frac{1}{\sqrt{2}} \begin{pmatrix}
0 & -i & 0 \\
i & 0 & -i \\
0 & i & 0
\end{pmatrix}, \quad \text{and} \quad  \hat{l}_z^\text{axi} =  \begin{pmatrix}
1 & 0 & 0 \\
0 & 0 & 0 \\
0 & 0 & -1
\end{pmatrix},
\end{equation}
with raising and lowering operators defined as $\hat{l}_\pm = \hat{l}_x\pm i \hat{l}_y.$ Together, these satisfy the algebra $[\hat{l}_z, \hat{l}_\pm] = \pm \hat{l}_\pm$ and $[\hat{l}_+,\hat{l}_-] = 2 \hat{l}_z.$ In the Cartesian (adjoint) representation, these operators take a non-diagonal form as
\begin{equation}
\hat{l}_x^\text{adj} = i \begin{pmatrix}
0 & 0 & 0 \\
0 & 0 & -1 \\
0 & 1 & 0
\end{pmatrix}, \quad \hat{l}_y^\text{adj} = i \begin{pmatrix}
0 & 0 & 1 \\
0 & 0 & 0 \\
-1 & 0 & 0
\end{pmatrix}, \quad \text{and} \quad  \hat{l}_z^\text{adj} =  i\begin{pmatrix}
0 & -1 & 0 \\
1 & 0 & 0 \\
0 & 0 & 0
\end{pmatrix}.
\end{equation} 
These two representations are related to each other by the unitary transformation $\mathcal{O}^\text{axi} = \mathcal{U}^\dagger \mathcal{O}^\text{adj} \mathcal{U},$ where
\begin{equation}
\mathcal{U} = \begin{pmatrix}
-\frac{1}{\sqrt{2}} & 0 & \frac{1}{\sqrt{2}} \\
-\frac{i}{\sqrt{2}} & 0 & -\frac{i}{\sqrt{2}} \\
0 & 1 & 0
\end{pmatrix}.
\end{equation}
In either representation, the Bloch Hamiltonian at each crystal momentum $\mathbf{k} = (k_x,k_y)$ can be partitioned into these operators
\begin{equation}
\label{eq: unperturbed Hamil}
\hat{\mathcal{H}}(\mathbf{k}) = h_0({\bf k}) \, \hat{\mathbb{I}} + h_c({\bf k}) \, \hat l_z \cdot \hat l_z + {\bf{h}}({\bf k}) \cdot \hat{\mathbb{L}},
\end{equation}
where  $h_0(\mathbf{k})$ is a scalar coupling, $h_c(\mathbf{k})$ is a crystal field, ${\bf h}({\bf k})= \left(h_1(\mathbf{k}), h_2(\mathbf{k}) \right)$ is a vector coupling to the orbital degree of freedom, $ \hat{\mathbb{L}}= \left( [\hat{l}_x \cdot \hat{l}_x- \hat{l}_y \cdot \hat{l}_y], [\hat{l}_x \cdot \hat{l}_y+ \hat{l}_y \cdot \hat{l}_x]\right),$
\begin{equation}
\begin{split}
h_0(\mathbf{k})&= t_\pi\bigg(\cos(ak_x)+2\cos\bigg(\dfrac{ak_x}{2}\bigg)\cos\bigg(\dfrac{\sqrt{3}ak_y}{2}\bigg)\bigg), \\
h_c(\mathbf{k})&= t_\sigma\bigg(\cos(ak_x)+2\cos\bigg(\dfrac{ak_x}{2}\bigg)\cos\bigg(\dfrac{\sqrt{3}ak_y}{2}\bigg)\bigg),\\
h_1(\mathbf{k})&= (t_\sigma-t_\pi)\bigg(\cos(ak_x)-\cos\bigg(\dfrac{ak_x}{2}\bigg)\cos\bigg(\dfrac{\sqrt{3}ak_y}{2}\bigg)\bigg), \\
h_2(\mathbf{k})&=-(t_\pi+t_\sigma)\sqrt{3}\sin\bigg(\dfrac{ak_x}{2}\bigg)\sin\bigg(\dfrac{\sqrt{3}ak_y}{2}\bigg),
\end{split}
\end{equation}
$a$ is the lattice constant, and $t_\pi$ and $t_\sigma$ are the hopping amplitudes of the $\pi$ and $\sigma$ bonds respectively. Note that $\hat{\mathbb{L}}$ is bilinear in angular momentum operators, and couples angular-momentum states differing by two units. In matrix form, the Hamiltonians can be written in the two representations as follows
\begin{equation}
\label{eq: adjoint}
\hat{\mathcal{H}}^\text{axi}(\mathbf{k}) = \begin{pmatrix}
h_0(\mathbf{k}) + h_c(\mathbf{k}) & 0 & h_1(\mathbf{k})- ih_2(\mathbf{k})\\
0 & h_0(\mathbf{k}) & 0 \\
h_1(\mathbf{k})+ ih_2(\mathbf{k}) & 0 & h_0(\mathbf{k}) + h_c(\mathbf{k})
\end{pmatrix},
\end{equation}
\begin{equation}
\label{eq: axial}
\hat{\mathcal{H}}^\text{adj}(\mathbf{k}) = \begin{pmatrix}
h_0(\mathbf{k}) + h_c(\mathbf{k}) - h_1(\mathbf{k}) & -h_2(\mathbf{k}) & 0 \\
-h_2(\mathbf{k}) & h_0(\mathbf{k}) + h_c(\mathbf{k})+ h_1(\mathbf{k}) & 0 \\
0 & 0 & h_0(\mathbf{k})
\end{pmatrix}.
\end{equation}

\section{B. Winding Numbers}
The sign selection for the winding numbers is determined by a competition of two energy scales, the bandwidth $\Delta$ determined by isotropic rotational invariants and intersite amplitudes $\delta$ . In an isotropic hopping model, the Bloch eigenstates have continuous $U(1)$ symmetry describing global in-plane orbital rotations and have twofold orbital degeneracies  that are removed by symmetry-allowed terms that couple the orbital polarizations to the lattice . The energy barrier $\Delta$ is the energy difference between the compensating point nodes at $K(K')$ and $\Gamma$, controlled by intersite rotational invariants of the form ${\mathbf{L}}_i \cdot {\mathbf{L}}_j$.  The second energy scale results from intersite amplitudes that couple the orbital degrees of freedom to the lattice ($\delta \sim {(\mathbf{L}}_i \cdot \hat{d}_{ij})( {\mathbf{L}}_j \cdot \hat{d}_{ij})$) everywhere except at the high-symmetry points where they are protected by discrete threefold ($K$ and $K'$) or sixfold ($\Gamma$) rotational symmetries.

\section{C. Anomalous Responses}

In order to activate an anomalous Hall response, we need to perturb the system with a $\mathcal{T}$-breaking potential. We consider a $\mathcal{T}$-breaking potential of the form
\begin{equation}
\hat{\lambda}_\mathcal{T}^\text{axi} = t_\mathcal{T} \begin{pmatrix}
1 & 0 & 0 \\
0 & 0 & 0 \\
0 & 0 & -1
\end{pmatrix},
\end{equation}
where $t_\mathcal{T}$ is the amplitude of the perturbation. This potential preserves $\mathcal{M}_z$-symmetry, but breaks $\mathcal{T}$-symmetry. Thus, it generates gaps at the  degeneracies in the mirror-even subspace at the $K,$ $K',$ and $\Gamma$ points, but preserves the line-node degeneracy between the mirror-odd and mirror-even subspaces. We can write this perturbation in the projected mirror-even subspace as $\varepsilon \sigma_z,$ with $\sigma_z$ being the diagonal Pauli matrix, as reported in the main text where $\varepsilon = t_\mathcal{T}$ in this case.

\begin{figure*}[htbp]
\centering
\includegraphics[width=6in]{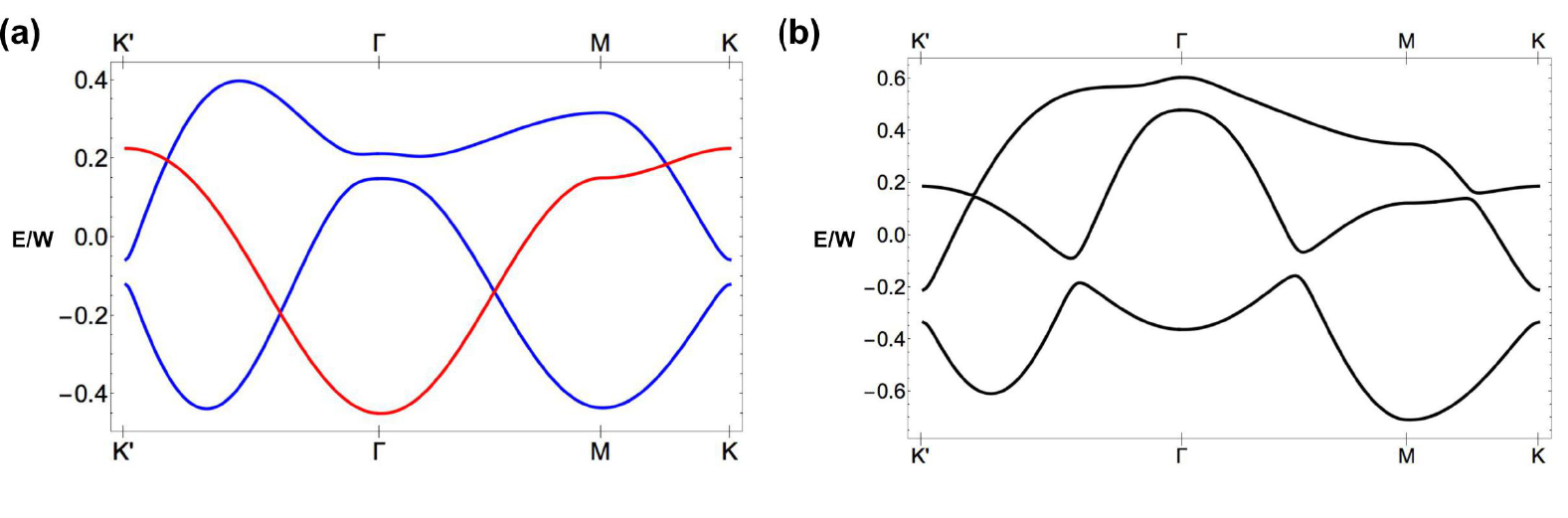}
\caption{\label{bandstructure} (a) Band structure associated with calculation of AHC given in Fig. 4a-b of the main text.  Mirror-even bands are shown in blue and the mirror-odd band in red.  Here, $t_\pi=-0.15$, $t_\sigma=0.21$, $t_\mathcal{T}=0.03$, and $t_{\mathcal{M}_z}=0$. (b) Band structure associated with calculation of AHC and AOHC given in Fig. 4c.  Here, $t_\pi=-0.12$, $t_\sigma=0.3$, $t_\mathcal{T}=0.063$, and $t_{\mathcal{M}_z}=0.00375$.}
\end{figure*}

On the other hand, to activate an anomalous \textit{orbital} Hall  response, we need to break both $z$-mirror symmetry and $\mathcal{T}$-symmetry. Such a mirror-breaking perturbation mixes the mirror-even and -odd subspaces, and thus, it is necessary to consider the full three-band model. We consider a $z$-mirror-breaking perturbation of the form
\begin{equation}
\hat{\lambda}_{\mathcal{M}_z}^\text{axi} = t_{\mathcal{M}_z} \begin{pmatrix}
0 & -1+i & 0 \\
-1-i & 0 & 1-i \\
0 & 1+i & 0
\end{pmatrix},
\end{equation}
where $t_{\mathcal{M}_z}$ is the magnitude of the perturbation. This perturbation gaps the line node between the even and odd subspaces into two degenerate points located off of the high-symmetry lines in the Brillouin zone.  Shown in Fig.~\ref{bandstructure} are the band structures used to calculate the AHC and AOHC reported in Fig. 4 of the main text.

In Fig. 1c of the main text, for the dashed lines, the parameters used for simulation are $t_\pi = -0.11$ and $t_\sigma = 0.33.$ For the solid lines, the parameters are $t_\pi = -0.11,$ $t_\sigma = 0.33,$ and $t_\mathcal{T} = 0.1.$ In Fig. 3 of the main text, the parameters used to simulate (a) are  $t_\pi=1$, $t_\sigma=0,$ and $\mu= -1.5,$ and those for (b) are $t_\pi= -0.1,$ $t_\sigma=0.2,$ $t_{M_z}=0.2$, $t_\mathcal{T}=0.3,$ and  $\mu= -0.18.$

\section{D. Coherent Optical Control}

To generate gaps in the energy spectrum, we couple the lattice coherently to a circularly-polarized optical field. In the long-wavelength limit, this interaction enters the Hamiltonian as the dipole interaction potential $\hat{\delta\mathcal{H}}(t) = -e \mathbf{r} \cdot \mathbf{E}(t),$ where $\mathbf{E}(t)$ is approximately position-independent. We consider an optical field propagating in the $z$-direction with vector amplitude $\mathbf{E} = E_0\mathbf{n},$ where $\mathbf{n} = (\cos \theta, \sin \theta)$ is a unit vector in the lattice plane, and $E_0$ is the field amplitude. The time-dependent field with frequency $\omega$ is
\begin{equation}
\label{eq: e field}
\mathbf{E}(t) = \mathbf{E} \cos \left( \omega t\right) +  \hat{z} \times \mathbf{E} \sin \left(\omega t \right).
\end{equation}
Here, the caret symbol on $x,y,z$ denotes unit vector. To obtain different polarizations, we negate the frequency $\omega \mapsto - \omega.$ In terms of this field, the interaction Hamiltonian is
\begin{equation}
\hat{\delta\mathcal{H}}(t) = -\frac{e}{2} \mathbf{r} \cdot \left( \boldsymbol{\mathcal{E}}_+ e^{i\omega t} + \boldsymbol{\mathcal{E}}_- e^{-i\omega t}\right),
\end{equation}
where we have isolated the positive- and negative-frequency components
\begin{equation}
\boldsymbol{\mathcal{E}}_\pm = \mathbf{E} \mp i \hat{z} \times \mathbf{E}.
\end{equation}
Without loss of generality, let us consider $\mathbf{E} = E_0 \hat{x};$ the interaction Hamiltonian simplifies to
\begin{equation}
\hat{\delta\mathcal{H}}(t)= -\frac{eE_0}{2}\left[ \left( x-iy \right) e^{i\omega t} + \left( x+iy \right) e^{-i\omega t}\right].
\end{equation}
Writing in the angular-momentum basis, the position operators can be represented as
\begin{equation}
x \pm i y = \mathfrak{p} \hat{l}_\pm.
\end{equation}
The interaction Hamiltonian is now
\begin{equation}
\hat{\delta\mathcal{H}}(t) = -\frac{eE_0\mathfrak{p}}{2}\left[ \hat{l}_- e^{i\omega t} + \hat{l}_+e^{-i\omega t}\right].
\end{equation}
In this form, the Hamiltonian manifestly conserves angular momentum. For every unit of angular momentum depleted from or absorbed by the optical field, a corresponding unit of angular momentum is added to or subtracted from the lattice respectively. Because of this, the optical field can hybridize the mirror-even and mirror-odd sectors of the band structure by balancing the angular momentum of the two subspaces.

We now study the effect of this interaction on the band structure. The unperturbed Hamiltonian $\hat{\mathcal{H}}$ is defined by Eq. ~\ref{eq: unperturbed Hamil}, and the full Hamiltonian $\hat{\mathcal{H}}'$ is now a function of $\mathbf{k}$ and $t,$ $\hat{\mathcal{H}}'(\mathbf{k},t) = \hat{\mathcal{H}}(\mathbf{k})+ \hat{\delta \mathcal{H}}(t).$ The time dependence is periodic with period $T = 2 \pi / \omega,$ so we can study this using the Floquet formalism \cite{Novi2017}. We can write the time-dependent Hamiltonian as an equivalent time-\textit{in}dependent Floquet Hamiltonian upon a Fourier transformation in time ($\hbar = 1$)
\begin{equation}
\hat{\mathcal{H}}_F(\mathbf{k}) = \tilde{\mathcal{H}}(\mathbf{k}) - \omega \mathcal{N},
\end{equation}
where $\mathcal{N}$ and $\tilde{\mathcal{H}}$ are defined by the following matrix elements
\begin{equation}
\begin{split}
\mathcal{N}_{m,n} &= m \delta_{m,n},\\
\tilde{\mathcal{H}}(\mathbf{k})_{m,n} & = \frac{1}{T} \int_0^T \hat{\mathcal{H}}'(\mathbf{k},t) e^{-i(n-m)\omega t}dt.
\end{split}
\end{equation}
Let us now analyze the zero-photon sector. By conservation of angular momentum, the zero-photon sector can only couple to the $n = \pm 1$ sectors. Therefore, the truncated Hamiltonian relevant to the zero-photon sector is
\begin{equation}
\hat{\mathcal{H}}_\text{0-photon} = \begin{pmatrix}
\hat{\mathcal{H}}(\mathbf{k})+\omega &  \Omega \hat{l}_- & 0 \\
 \Omega \hat{l}_+ & \hat{\mathcal{H}}(\mathbf{k}) &  \Omega \hat{l}_- \\
0 &  \Omega \hat{l}_+ & \hat{\mathcal{H}}(\mathbf{k})-  \omega
\end{pmatrix},
\end{equation}
where $\Omega = -eE_0 \mathfrak{p}/2.$ We further project to the mirror-even subspace of the zero-photon sector to obtain an effective Hamiltonian, in the axial representation,
\begin{equation}
\begin{split}
\hat{\mathcal{H}}_\text{eff}(\mathbf{k}) &= \hat{\mathbb{P}} \mathcal{H}(\mathbf{k})\hat{\mathbb{P}}+ \frac{\Omega^2}{h_c(\mathbf{k})-\omega} \hat{\mathbb{P}} \hat{l}_+ \hat{\mathbb{Q}} \hat{l}_- \hat{\mathbb{P}}+ \frac{\Omega^2}{h_c(\mathbf{k})+\omega} \hat{\mathbb{P}} \hat{l}_- \hat{\mathbb{Q}} \hat{l}_+ \hat{\mathbb{P}} = (h_0(\mathbf{k})+h_c(\mathbf{k}))\hat{\mathbb{I}}+ h_1(\mathbf{k}) \sigma_x+h_2(\mathbf{k}) \sigma_y + \varepsilon(\mathbf{k}) \sigma_z,
\end{split}
\end{equation}
where  $\hat{\mathbb{P}}$ and $\hat{\mathbb{Q}} = \hat{\mathbb{I}}-\hat{\mathbb{P}}$ are projection operators, and the induced gap is given by
\begin{equation}
\varepsilon(\mathbf{k}) = \frac{2 \Omega^2 \omega}{h_c(\mathbf{k})^2-\omega^2} = \frac{e^2E_0^2 \mathfrak{p}^2 \omega}{2(h_c(\mathbf{k})^2-\omega^2)}.
\end{equation}

\end{document}